\documentclass[11pt]{article}
\usepackage[margin=1in]{geometry}
\usepackage{graphicx}
\usepackage{amsmath}

\def\beq{\begin{equation}}
\def\eeq#1{\label{#1}\end{equation}}
\def\eeqn{\end{equation}}

\def\beqa{\begin{eqnarray}}
\def\eeqa#1{\label{#1}\end{eqnarray}}
\def\eeqan{\end{eqnarray}}

\let\bar=\overbar

\def\etal{{\it et al.}}

\def\Dslash{\not{\hbox{\kern-4pt $D$}}}
\def\dslash{\not{\hbox{\kern-2pt $\del$}}}

\def\msb{{\bar{\ssstyle M \kern -1pt S}}}

\def\Title#1{\begin{center} {\Large {\bf #1} } \end{center}}
\def\Author#1{\begin{center} {\normalsize {\sc #1} } \end{center}}
\def\Institution#1{\begin{center} {\normalsize {\it #1} } \end{center}}
\def\Abstract#1{\noindent {\normalsize {\bf Abstract:} {\normalfont #1}}}
\def\Conference{\vspace{4mm}\begin{raggedright} {\normalsize {\it Talk presented at the 2019 Meeting of the Division of Particles and Fields of the American Physical Society (DPF2019), July 29--August 2, 2019, Northeastern University, Boston, C1907293.} } \end{raggedright}\vspace{4mm}}

\begin{document}

%
%

\Title{Searching for Lorentz violation in high-energy colliders}

\Author{Zonghao Li}

\Institution{Department of Physics and IU Center for Spacetime Symmetries
\\ Indiana University, Bloomington, IN 47405, USA}

\Abstract{
Lorentz violation has been a popular field in recent years 
in the search for new physics beyond the Standard Model. 
We present a general method to build all Lorentz-violating terms in gauge field theories, 
including ones involving operators of arbitrary mass dimension. 
Applying these results to two types of experiments in high-energy colliders, 
light-by-light scattering and deep-inelastic scattering, 
we extract first bounds on certain coefficients for Lorentz violation.}

\Conference

%
%

\section{Introduction}

Interest in the study of Lorentz violation has been growing
in recent years.
Lorentz symmetry as a fundamental symmetry
in the Standard Model (SM) and General Relativity (GR)
deserves to be tested precisely in experiments.
Moreover,
the unification of the SM and GR at a high-energy level
should leave new signals beyond conventional physics
in low-energy experiments.
Lorentz violation is a popular candidate among those new signals.
Testing of Lorentz violation can shed light 
on the underlying theory unifying the SM and GR.

Experiments in high-energy colliders played a crucial role
in the building of the SM.
Now,
they are also universally applied in the search for new physics beyond the SM.
Searching for Lorentz violation in high-energy colliders
can provide us fruitful insight into this new physics.
To describe Lorentz violation in these experiments,
we adopt the Standard-Model Extension (SME) framework
developed by D.\ Colladay and V.A.\ Kosteleck\'y 
\cite{ck97}.
The SME is a comprehensive framework 
that can describe all possible Lorentz-violating signals
in the context of effective field theory.

In this work,
we develop a general method to build all possible Lorentz-violating terms
in gauge field theories.
Applying this to quantum electrodynamics (QED) and quantum chromodynamics (QCD),
we study two experiments,
light-by-light scattering and deep-inelastic scattering,
to search for Lorentz violation in high-energy colliders.
This contribution to the DPF2019 proceedings is based on Ref.\ 
\cite{kl19}.
The derivations are performed in flat spacetime,
but related techniques can also be applied in gravity
\cite{kl20}.

\section{Theoretical framework}
 
To describe all possible Lorentz-violating signals in experiments,
the SME includes all possible Lorentz-violating corrections to the Lagrange density.
All minimal terms
\cite{ck97,ak04}
and nonminimal free-propagation terms 
\cite{km09}
in the SME have been established,
where the term minimal refers to operators of mass dimension $d\leq4$.
Nonminimal interaction terms with low mass dimension $d\leq6$ in QED
have also been established
\cite{dk16}.
However,
the general form of Lorentz-violating terms with arbitrary mass dimension is still unknown.

In the work summarized here,
we developed a general method to build all possible Lorentz-violating terms 
in gauge field theories 
\cite{kl19}.
Since the SME preserves gauge invariance,
the key point is to find all possible gauge-invariant operators.
We build gauge-invariant operators through gauge-covariant operators,
which are combinations of gauge-covariant derivatives
and gauge strength tensors.
All these kinds of operators are completely classified and enumerated 
into linear combinations of a standard basis set.
The general Lorentz-violating terms are constructed 
by combining gauge-invariant operators with coefficients for Lorentz violation,
which are fixed in the vacuum.
Applying these in QED and QCD,
we build the full Lorentz-violating extensions of QED and QCD
that can be tested in high-energy colliders.
Two examples,
light-by-light scattering and deep-inelastic scattering,
are explored in detail in the following sections
to look for Lorentz violations in QED and QCD, 
respectively.

\section{Sidereal-time dependence}

Before turning to specific experiments,
we first discuss an important method
to search for Lorentz violation in experiments:
sidereal-time dependence.
The coefficients for Lorentz violation are fixed in the vacuum,
but the Earth rotates about its own axis and revolves around the Sun.
As a result,
experimental results depend on sidereal time
in presence of Lorentz violation.
To show this quantitatively,
we introduce the Sun-centered frame 
\cite{datatables},
which is close to an inertial frame.
We assume coefficients for Lorentz violation
to be approximately constant in the Sun-centered frame. 

The origin of $T$ axis of the Sun-centered frame
is set as the 2000 vernal equinox.
The $Z$ axis aligns with the Earth axis.
The $X$ axis points from the Sun to the Earth at the 2000 vernal equinox,
and the $Y$ axis forms a right-hand system with the $X$ and $Z$ axes.
The boost of the Earth in the Sun-centered frame,
which is around $10^{-4}$,
can be omitted in analyses of most experiments.
Then,
the transition between the laboratory frame and the Sun-centered frame
can be expressed as a rotation matrix:
\beq
R^{jJ}=\left(
\begin{array}{ccc}
\cos\chi \cos\omega_\oplus T_\oplus &
\cos\chi \sin\omega_\oplus T_\oplus &
-\sin\chi \\
-\sin\omega_\oplus T_\oplus &
\cos\omega_\oplus T_\oplus &
0\\
\sin\chi \cos\omega_\oplus T_\oplus &
\sin\chi \sin\omega_\oplus T_\oplus &
\cos\chi
\end{array}
\right),
\eeq{eq:sun}
where $\omega_\oplus \simeq 2\pi/(23$ h $56$ min$)$
is the Earth's sidereal rotation frequency,
$T_\oplus$ is the local sidereal time,
$\chi$ is the angle between the $z$ axis in the local laboratory frame
and the $Z$ axis in the Sun-centered frame,
$j=x,y,z$ are spatial coordinate indices in the local laboratory frame,
and $J=X,Y,Z$ are spatial coordinate indices in the Sun-centered frame.

In an experimental analysis,
we can decompose the sidereal dependence
into different orders of harmonics in sidereal time.
Different coefficients for Lorentz violation may contribute
to different orders of harmonics.
In this way,
we can distinguish contributions from different coefficients for Lorentz violation.
In summary,
sidereal dependence is the most crucial signal of Lorentz violation in experiments.
Any nontrivial observation of sidereal dependence 
would be a clear evidence for Lorentz violation
and imply new physics beyond the SM.

\section{Light-by-light scattering}

Precision tests of QED provided strong evidence for modern physics 
and continue shedding light on possible new physics beyond the SM.
One of the latest tests of QED is light-by-light scattering studied at the LHC 
\cite{aetal17}.
Light-by-light scattering is absent in classical electrodynamics
because of the linearity of the Maxwell equations.
It is instead a pure quantum effect of QED from one-loop radiative correction.
Therefore,
testing light-by-light scattering is a direct test of quantum effects from QED.
Due to experimental difficulties in preparing high-density and high-energy photon flux,
direct tests of light-by-light scattering
have not been performed until recently at the LHC 
\cite{aetal17}.
The recent experiment used ultraperipheral Pb+Pb collisions 
at center-of-mass energy $\sqrt{s_{NN}}=5.02$ TeV per nucleon pair.
By the equivalent photon approximation 
\cite{epa},
this can be viewed as collisions of photons.

\begin{figure}[htb]
\centering
\includegraphics[height=2in]{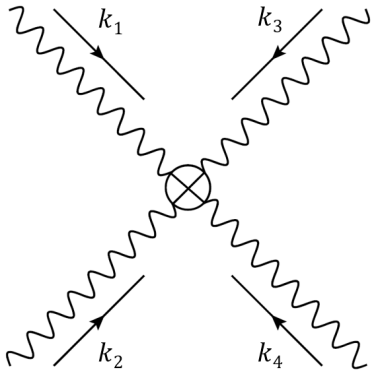}
\caption{Dominant contribution to light-by-light scattering.}
\label{fig:photon}
\end{figure}

Light-by-light scattering can be used to test Lorentz symmetry.
In fact,
the SME as a framework in the context of effective field theory 
can describe all possible deviations from SM predictions.
The leading-order contribution of the SME 
comes from a term with mass dimension $d=8$ in Lagrange density:
\beq
\mathcal{L}_g^{(8)}\supset -\tfrac{1}{48} 
k_F^{(8)\kappa\lambda\mu\nu\rho\sigma\tau\upsilon}
F_{\kappa\lambda}F_{\mu\nu}F_{\rho\sigma}F_{\tau\upsilon},
\eeq{eq:light}
where $k_F^{(8)\kappa\lambda\mu\nu\rho\sigma\tau\upsilon}$
are coefficients for Lorentz violation for mass dimension $d=8$.
They are components of a rank-eight tensor field
and are fixed in the vacuum.
They generally give preferred directions in the vacuum,
which would cause violation of Lorentz symmetry.
The above term includes four electromagnetic tensors,
so it introduces a new interaction vertex with four photon lines
as shown in Fig.\ \ref{fig:photon}.

The Lagrange density in Eq.\ (\ref{eq:light})
creates many new experimental signals for light-by-light scattering.
It gives new contributions to the total cross section
and also alters the differential cross section.
Since the coefficients for Lorentz violation are fixed in the vacuum
and are assumed approximately constant in the Sun-centered frame,
the experimental result can depend on the sidereal time 
as discussed in the last section.
The experimental result can also depend on the location and velocity of the laboratory
due to Lorentz violation.
The energy dependence of the cross section
is also modified.

Due to limitations of the existing experimental data,
we only study the total cross section here.
The LHC experiment measured the total cross section of light-by-light scattering
to be $70\pm24$(stat.)$\pm17$(syst.) nb 
\cite{aetal17}.
Meanwhile,
the theoretical prediction from the SM is $49\pm10$ nb
\cite{ks10}.
Comparing these two results,
we can extract bounds on the $k_F^{(8)}$-type coefficients.
We also get constraints on the Lorentz-invariant and isotropic
combinations of the $k_F^{(8)}$-type coefficients.
They are all bounded around $10^{-7}$ GeV$^{-4}$
\cite{kl19}.
More precise measurements of the cross section
would improve these results.
Experiments can also measure the sidereal dependence of the cross section,
which would provide clear evidence of Lorentz violation.

\section{Deep-inelastic scattering}

Deep-inelastic scattering provides key experimental support for QCD and the quark model.
It also serves as an essential tool in the search for new physics.
The SME framework as an effective field theory
can describe all possible deviations from the SM predictions.

Previous work studied the effects from minimal terms in the SME
\cite{klv17}.
Since we constructed the general terms in QCD and QED,
the contributions from nonminimal terms can also be studied.
The leading-order spin-independent contribution in the nonminimal SME
comes from a mass-dimension $d=5$ term:
\beq
\mathcal{L}_\psi^{(5)}\supset
-\tfrac12\sum_f
a_f^{(5)\mu\alpha\beta}\overline{\psi}_f\gamma_\mu 
iD_{(\alpha}iD_{\beta)} \psi_f+\textrm{h.c.},
\eeq{eq:DIS}
where $a_f^{(5)\mu\alpha\beta}$ are coefficients for Lorentz violation,
$D$ is the gauge-covariant derivative,
$f=u,d$ runs over the dominant quark flavors,
parentheses in lower indices 
mean symmetrization over $\alpha$ and $\beta$ with a coefficient $1/2$,
and h.c.\ means hermitian conjugation.
We only considered spin-independent contributions here
because most experiments use unpolarized beams.
As before,
the $a^{(5)}$-type coefficients are fixed in vacuum
and generally violate Lorentz symmetry.

The Lagrange density in Eq.\ (\ref{eq:DIS})
changes both free propagations and interactions of quarks.
The parton model is also slightly modified.
Taking all these corrections into account,
we can get the modified differential cross section
after some calculations.
The calculation details and results are presented in Ref.\ 
\cite{kl19}.

Based on the simulations on $c$-type coefficients in Ref.\
\cite{klv17},
we estimate the nonminimal $a^{(5)}$-type coefficients 
could be bounded around $10^{-7}-10^{-4}$ GeV$^{-1}$.
The $a^{(5)}$-type coefficients 
also introduce many new signals 
that are absent if only $c$-type coefficients are nonzero.
Since the $a^{(5)}$-type coefficients
contain three spacetime indices,
the cross section can depend on up to the third-order harmonics
of sidereal time.
Moreover,
$a^{(5)}$-type coefficients are CPT odd,
so they contribute differently to scatterings with proton and antiproton.
Tests of these new signals would be very helpful 
in the search for new physics beyond the SM.



\section*{Acknowledgments}

This work was supported in part by the U.S. Department of Energy
and by the Indiana University Center for Spacetime Symmetries.

\end{document}